\documentclass[12pt,preprint,aps,english,superscriptaddress]{revtex4-1}
\usepackage[T1]{fontenc}
\usepackage[latin9]{inputenc}
\usepackage{amsmath}
\usepackage{amssymb}
\usepackage{cancel}
\usepackage{graphicx}
\usepackage{esint}
\usepackage{color}
\usepackage[version=4]{mhchem}
\usepackage{siunitx, xspace, bm, changes}
\makeatletter

\newcommand{\bk}{{\bm k}}
\newcommand{\SRO}{\ce{Sr2RuO4}}

\@ifundefined{textcolor}{}{%
	\definecolor{BLACK}{gray}{0}
	\definecolor{WHITE}{gray}{1}
	\definecolor{RED}{rgb}{1,0,0}
	\definecolor{GREEN}{rgb}{0,1,0}
	\definecolor{BLUE}{rgb}{0,0,1}		
	\definecolor{CYAN}{cmyk}{1,0,0,0}
	\definecolor{MAGENTA}{cmyk}{0,1,0,0}
	\definecolor{YELLOW}{cmyk}{0,0,1,0}
}

\usepackage{dcolumn}% Align table columns on decimal point
\usepackage{bm}% bold math
\usepackage{enumerate}

\usepackage[caption=false]{subfig}

\@ifundefined{showcaptionsetup}{}{%
	\PassOptionsToPackage{caption=false}{subfig}}
\usepackage{babel}
\addto\captionsenglish{}
\renewcommand{\thefigure}{\@arabic\c@figure}

\makeatother

\begin{document}
	
\title{Hall coefficient signals orbital differentiation\\ in the Hund's metal Sr$_2$RuO$_4$}

\author{M. Zingl}
\email[]{mzingl@flatironinstitute.org}
\affiliation{Center for Computational Quantum Physics, Flatiron Institute, 162 5th Avenue, New York, NY 10010, USA}
\author{J. Mravlje}
\affiliation{Jo\v zef Stefan Institute, Jamova 39, Ljubljana, Slovenia}
\author{M. Aichhorn}
\affiliation{Institute of Theoretical and Computational Physics, Graz University of Technology, NAWI Graz, 8010 Graz, Austria}
\author{O. Parcollet}
\affiliation{Center for Computational Quantum Physics, Flatiron Institute, 162 5th Avenue, New York, NY 10010, USA}
\affiliation{Institut de Physique Th\'{e}orique (IPhT), CEA, CNRS, UMR 3681, 91191 Gif-sur-Yvette, France}
\author{A. Georges}
\affiliation{Coll\`ege de France, 11 place Marcelin Berthelot, 75005 Paris, France}
\affiliation{Center for Computational Quantum Physics, Flatiron Institute, 162 5th Avenue, New York, NY 10010, USA}
\affiliation{Centre de Physique Th\'{e}orique Ecole Polytechnique, CNRS, Universit\'e Paris-Saclay, 91128 Palaiseau, France}
\affiliation{Department of Quantum Matter Physics, University of Geneva, 24 Quai Ernest-Ansermet, 1211 Geneva 4, Switzerland}
\date{\today}

\begin{abstract}
\textbf{
The Hall coefficient $R_H$ of \SRO{} exhibits a non-monotonic temperature dependence  with two sign reversals. 
We show that this puzzling behavior is the signature of two crossovers which are key to the physics of this material. The increase of $R_H$ and the first sign change upon cooling are 
associated with a crossover into a regime of coherent quasiparticles with strong orbital differentiation of the
inelastic scattering rates. The eventual decrease and the second sign change at lower temperature is driven by the crossover from inelastic to impurity-dominated scattering. 
This qualitative picture is supported by quantitative calculations of $R_H(T)$ using
Boltzmann transport theory in combination with dynamical mean-field theory, taking into account
the effect of spin-orbit coupling. Our insights shed new light on the temperature dependence of the Hall coefficient
in materials with strong orbital differentiation, as observed in Hund's metals.
}
\end{abstract}

\maketitle

\section*{Introduction}
Measuring the Hall coefficient $R_H$ is a standard way of characterizing charge carriers in quantum materials.
For free carriers of a single type the Hall coefficient $R_H$ is simply
given by the inverse of the density of carriers $n$ and their charge $e$. However, in complex
materials with a Fermi surface (FS) composed of multiple sheets, interpreting $R_H$ can be more
complicated and also provides richer information when both electron-like and hole-like carriers are present. For instance, in the
case of one hole-like and one electron-like FS sheet, the corresponding Hall coefficient is given
by an average of $R_{H,e}<0$ and $R_{H,h}>0$:
\begin{equation}
\label{eqn:RH_average}
R_H = \frac{\sigma^2_e \, R_{H,e}  + \sigma^2_h \, R_{H,h}}{\left(\sigma_e + \sigma_h\right)^2}\, ,
\end{equation}
weighted by the squares of the individual hole and electron conductivities, $\sigma_h$ and
$\sigma_e$ respectively. Hence, the ratio of scattering rates between the two types of carriers
enters in a key manner to determine both the overall sign and magnitude of the Hall coefficient.

\begin{figure}[h]
	\includegraphics[width=1.0\linewidth]{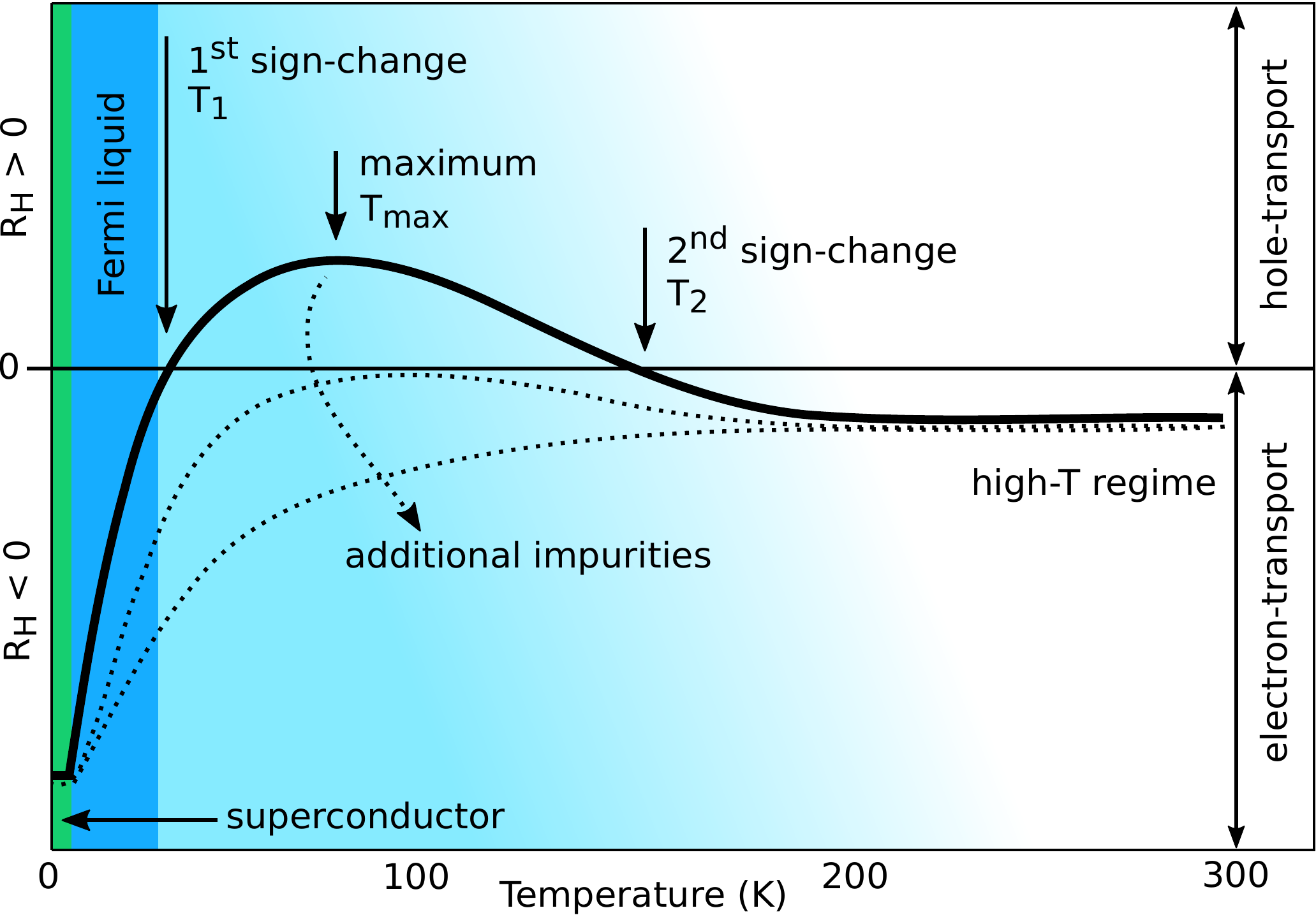}
	\caption{Sketch of the temperature dependence of the Hall coefficient $R_H$ (solid line) and
		the different transport/electronic regimes in \SRO{} after experimental data from
		Refs.~\cite{Shirakawa1995, Mackenzie1996, Galvin2001}. $R_H$ changes sign twice at about
		\num{30} and \SI{120}{K}, which can be suppressed by adding small amounts of \ce{Al}
		impurities (dashed lines)~\cite{Galvin2001}.}
	\label{fig:fig1}
\end{figure}

The 4d transition metal oxide \SRO{} is such a complex material: with low-energy bands built
out of three Ru-$t_{2g}$ orbitals ($d_{xy},d_{yz},d_{xz}$) hybridized with O-$2p$ states, it
has a FS comprising two electron-like sheets, $\beta$ and $\gamma$, and one hole-pocket,
$\alpha$~\cite{Mackenzie1996b, Bergemann2000,Damascelli2000}. And indeed,
experiments~\cite{Shirakawa1995,Mackenzie1996,Galvin2001} have observed a particularly
intriguing temperature dependence of $R_H$ in \SRO{}, as
depicted on Fig.~\ref{fig:fig1}. $R_H$ increases
from a negative value of about \SI{-1 e-10}{\cubic\meter\per\coulomb} at low temperatures (values between \num{-1.37 e-10} and \SI{-0.7 e-10}{\cubic\meter\per\coulomb}
for $T\rightarrow 0$ have been reported~\cite{Shirakawa1995,Mackenzie1996,Galvin2001,Kikugawa2004}),
exhibits a sign change at $T_1 = \SI{30}{K}$ (in the cleanest samples), reaches a positive maximum at about
\SI{80}{K}, changes sign a second time around $T_2 = \SI{120}{K}$ and eventually saturates to a slightly
negative value for $T > \SI{200}{K}$.

Obviously, this rich temperature dependence is a signature of the multi-carrier nature of \SRO, 
as realized early on in Ref.~\cite{Shirakawa1995}, which considered a Drude model with two types
of carriers. This was refined later in several works~\cite{Mazin2000, Noce2000, Noce1999} using
Boltzmann transport theory calculations for tight-binding models assuming scattering rates
$1/\tau_\nu = A_\nu + B_\nu\,T^2$ for the different FS sheets $\nu = \{\alpha, \beta, \gamma \}$,
with adjustable parameters $A_\nu$ and $B_\nu$. The overall take-home message of these
phenomenological models is that $R_H$ is highly sensitive to the precise details of the FS sheets
and also to the temperature and sheet dependence of the scattering rates.

Another remarkable experimental finding provides insight in interpreting the temperature dependence
of $R_H$~\cite{Galvin2001}: Adding small amounts of Al impurities has a drastic impact on the
intermediate temperature regime such that $R_H$ no longer turns positive and instead increases
monotonically from the low-T to the high-T limit, as indicated by dotted lines in Fig.~\ref{fig:fig1}.
Arguably, the similarity of the low-T values of $R_H$ for different impurity concentrations
provides evidence that the elastic-scattering regime has been reached where $R_H$ is mainly determined
by FS properties (see also Ref.~\cite{Mackenzie1996}). In contrast, the temperature dependence
itself must be due to inelastic scattering, possibly associated with electronic correlations~\cite{Galvin2001}.

In this work, we address this rich temperature dependence of $R_H$ in \SRO{} and
provide a clear interpretation of its physical meaning.
We show that the two sign changes of $R_H(T)$ in clean samples are the signatures of two
important crossovers in the physics of this material. The increase of $R_H$ upon cooling from
high temperature signals the gradual formation of coherent quasiparticles, which is
associated with a strong temperature dependence of the ratio of inelastic scattering rates between
the $xy$ and $xz/yz$ orbitals. At low temperatures the decrease of $R_H$ is due to the crossover from 
inelastic to impurity-dominated scattering. These qualitative insights have relevance
to a wide class of materials with orbital differentiation.

Our qualitative picture is supported by a quantitative calculation of $R_H(T)$
using Boltzmann transport theory in combination with dynamical mean-field theory
(DMFT)~\cite{Georges1996}, taking into account the electronic structure of the material.
The spin-orbit coupling (SOC) is found to play a key role~\cite{Haverkort2008,Veenstra2014}, because
it has a strong influence on the shape of the FS and also controls the manner in which the
scattering rates associated with the different orbitals combine into $\bk$-dependent quasiparticle
scattering rates at a given point on the FS~\cite{Haverkort2008,Veenstra2014,Tamai2018}.

\section*{Results}
 
\emph{Dependence on scattering rate ratios}

The orbital dependence of scattering rates is crucial for the understanding of the Hall effect. Therefore, we begin
by assigning scattering rates $\eta_{xy}$, $\eta_{xz}$ and $\eta_{yz}$ (due to crystal symmetries $\eta_{xz}=\eta_{yz}$)
to each orbital, irrespective of the microscopic details of the underlying scattering mechanisms, which will be
addressed at a later stage. Then, these scattering rates are converted into band and $\bk$-dependent scattering rates using the overlap of the orbital wave-function with the Bloch wave-function:
\begin{equation}\label{eq:2}
\eta_\nu(\mathbf{k}) = \sum_{m} |\langle \psi_{\mathbf{k}}^\nu | \chi_m\rangle|^2 \ \eta_m \ \ \ \text{with} \ \ \  m = \{xy,xz,yz\}.
\end{equation}
We calculate $R_H$ with Boltzmann transport theory using a realistic Wannier-Hamiltonian for the effective
low-energy \ce{Ru}-$t_{2g}$ subspace (see Methods). 
In Boltzmann transport theory $R_H$ only depends on the scattering rates
through their ratio $\xi = \eta_{xy}/\eta_{xz/yz}$ and not their absolute magnitude; a point we verified in our
calculations. This also implies that within the constant isotropic scattering rate approximation,
i.e. $\xi = 1$, the full temperature dependence of $R_H$ cannot be explained.

\begin{figure}[h]
	\includegraphics[width=1.0\linewidth]{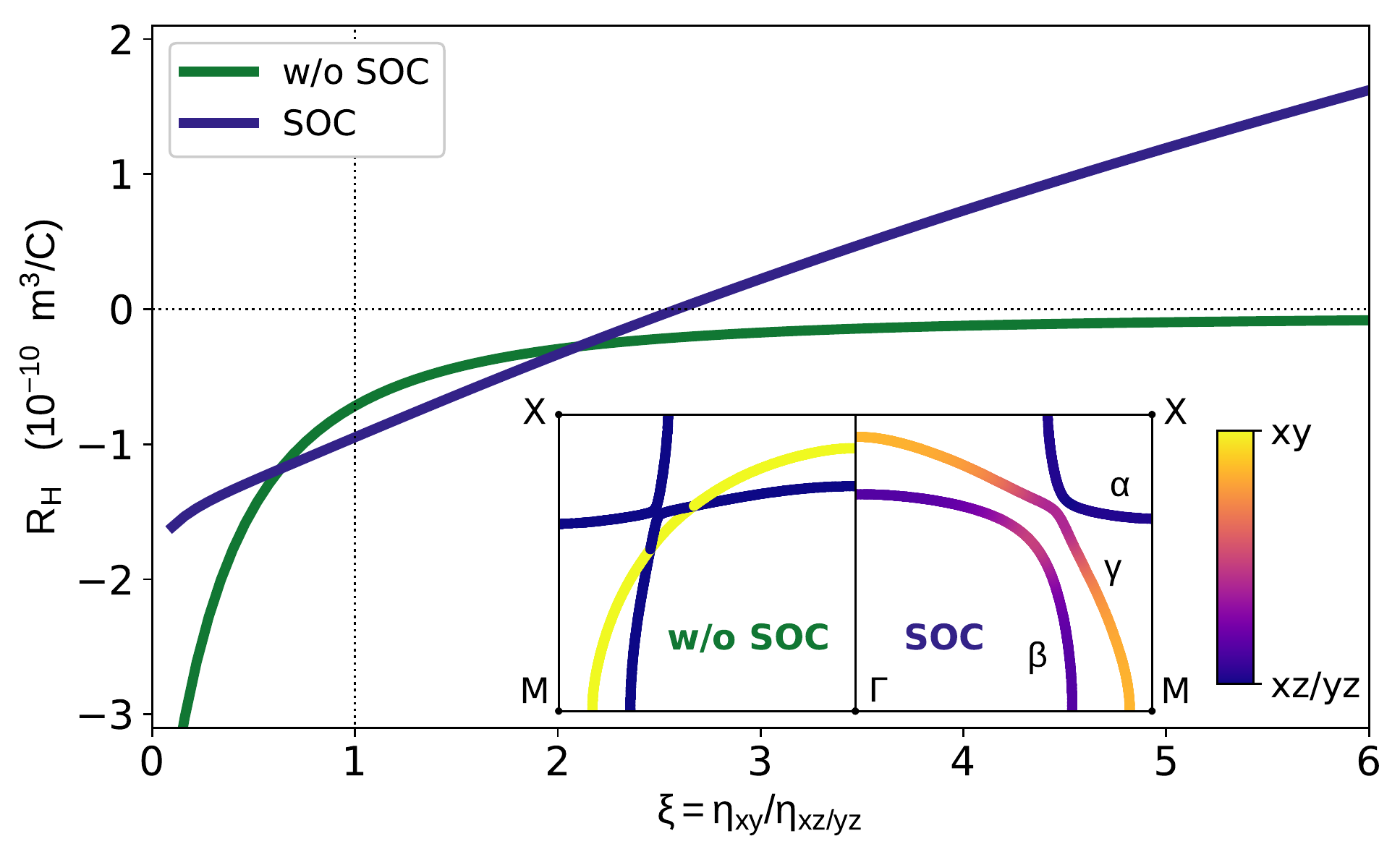}
	\caption{Dependence of the Hall coefficient $R_H$ on the ratio of scattering rates
		$\xi = \eta_{xy}/\eta_{xy/yz}$ with and without SOC. The inset shows the orbital character of the
		Fermi surface sheets and the influence of SOC on their shape for $k_z = 0$.}
	\label{fig:fig2}
\end{figure}

The calculated $R_H$ as a function of the scattering rate ratio $\xi$ is displayed in Fig.~\ref{fig:fig2}.
We compare calculations with and without SOC, and note that in the following all results labeled with `SOC'
actually take the correlation-enhancement of the  effective SOC into account~\cite{Liu2008,Zhang2016,Kim2018,Tamai2018} (see Methods).
Without SOC $R_H$ remains negative for all values of $\xi$ and approaches zero as $\xi \gg 1$. In this
limit the $\gamma$ sheet drops out and the contributions of the hole-like $\alpha$ sheet and
electron-like $\beta$ sheet compensate each other. This means that it is not possible to explain
the positive value of $R_H$ observed experimentally in clean samples for $T_1<T<T_2$ (Fig.~\ref{fig:fig1})
without taking SOC into account. With SOC we observe a very different behavior of $R_H(\xi)$; it turns
from negative to positive at $\xi \simeq 2.6$.
This is a result of two effects~\cite{Haverkort2008,Veenstra2014,Tamai2018} (see Fig.~\ref{fig:fig2}, inset):
First, SOC changes the shape and size of the FS sheets, and secondly, it induces a mixing between different
orbital characters, which varies for each point on the FS. Thus, the manner in which the scattering rates
associated with the different orbitals combine into $\bk$-dependent quasiparticle scattering rates (Eq.~\ref{eq:2})
is controlled by the SOC. From the calculated dependence of $R_H(\xi)$
in the presence of SOC we deduce that agreement with experiments would require $\xi$ to be smaller than
$2.6$ at high temperatures, increase above this value at $\sim T_2$ and then decrease again to reach a
value close to unity at low temperatures.

\emph{Inelastic electron-electron scattering}

\begin{figure}[h]
	\includegraphics[width=1.0\linewidth]{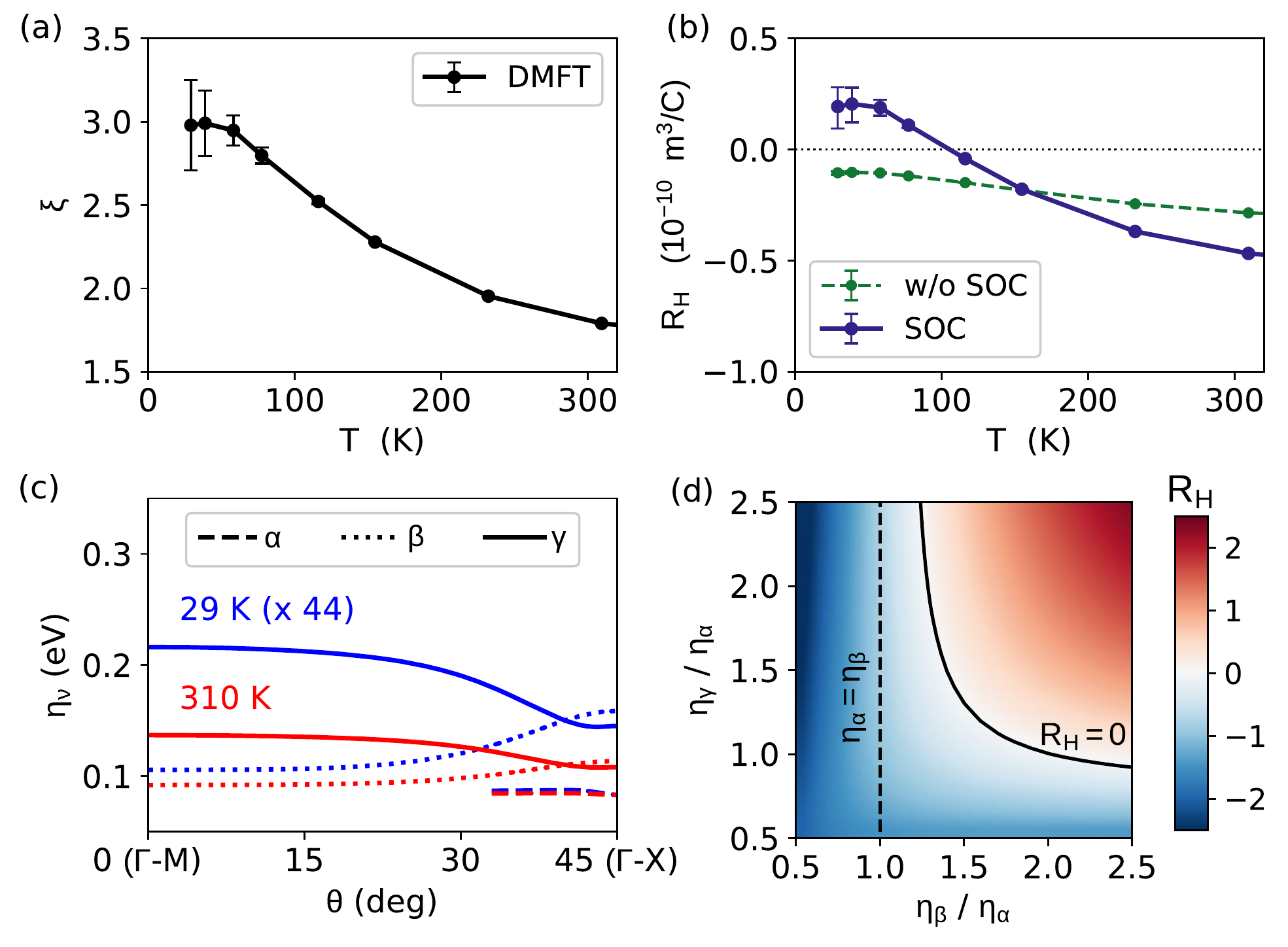}
	\caption{(a) Inelastic scattering rate ratios extracted from DMFT self-energies extrapolated to zero frequency (see Methods). The errorbars represent the standard deviation of 9 consecutive DMFT
		iterations. (b) Influence of inelastic electron-electron scattering on $R_H(T)$. (c) Scattering
		rates $\eta_{\nu}(\theta)$ on the three Fermi surface sheets $\nu=\{\alpha,\beta,\gamma\}$ for
		$k_z = 0$ along the angle $\theta$ from $0^\circ$ ($\Gamma$-M) to $45^\circ$ ($\Gamma$-X) under
		consideration of the orbital character shown in the right inset of Fig.~\ref{fig:fig2}. The scattering
		rates at \SI{29}{K} (blue) are multiplied by a factor of 44, chosen such that
		$\eta_\alpha\left(45^\circ\right)$ coincides with the result at \SI{310}{K} (red). (d) Color map
		of $R_H$ assuming constant sheet-dependent scattering rate ratios. The solid black line indicates
		$R_H = 0$ and the dashed black line marks $\eta_\alpha = \eta_\beta$.}
	\label{fig:fig3}
\end{figure}

We turn now to microscopic calculations by first considering inelastic electron-electron scattering ratios
calculated with DMFT (see Methods). These calculations consider the $t_{2g}$ subspace of states with
Hubbard-Kanamori interactions of $U=\SI{2.3}{eV}$ and $J=\SI{0.4}{eV}$~\cite{Mravlje2011}.
The calculated $\xi(T)$ from inelastic scattering only is displayed in Fig.~\ref{fig:fig3}~(a).
In agreement with previous
studies~\cite{Mravlje2011, Deng2016}, we find that the $xy$ orbital is less coherent than $xz/yz$ at
all temperatures and $\eta_{xy} > \eta_{xz/yz}$. In \SRO{} the crossover from the low-T coherent Fermi
liquid regime with $\eta \sim T^2$ to an incoherent regime with a quasilinear temperature dependence of
the scattering rate is well-documented~\cite{Mravlje2011, Stricker2014} and also manifested in deviations
of the resistivity from a low-temperature quadratic behavior to a linear one~\cite{Hussey1998}.
Importantly, this coherence-to-incoherence crossover as well as the corresponding coherence scales are
strongly orbital dependent. When approaching the Fermi liquid regime ($T_{FL} \approx \SI{25}{K}$~\cite{Hussey1998,Maeno1997,Mackenzie2003}) the scattering rate ratio reaches a value as large as $\xi^{FL} \sim 3$ (Fig.~\ref{fig:fig3}~(a)), but decreases rapidly upon heating with
$\xi = 1.8$ at \SI{300}{K}. We do not find a substantial change for even higher temperatures; at 500 K the scattering rate ratio is $\xi= 1.6$.

Connecting these results to the discussion of Fig.~\ref{fig:fig2} above, the temperature dependence of
$\xi$ directly translates into that of $R_H$, as shown in Fig.~\ref{fig:fig3}~(b). Like in experiments, $R_H$
is negative at high temperatures, but when the temperature is lowered it increases and crosses zero at
\SI{110}{K}. This demonstrates that electronic correlations are indeed able to turn $R_H$ positive and
suggests the following physical picture: The electronic transport in \SRO{} crosses from a regime governed
by incoherent electrons at high temperatures, connected to a weaker orbital differentiation of scattering rates
and a negative $R_H$, over to a coherent Fermi liquid regime, with a stronger orbital differentiation and
positive $R_H$. The resulting sign change at \SI{110}{K} can be seen as a direct consequence of this
coherence-to-incoherence crossover. We emphasize that this sign change is only observed when SOC is taken
into account. Without SOC $R_H$ is purely negative and shows only a weak temperature dependence
(Fig.~\ref{fig:fig3}~(b), dashed line).

When moving along the FS from $\Gamma$-M ($\theta=0^\circ$) to $\Gamma$-X ($\theta=45^\circ$), the mixing
of the orbital character induced by SOC (Eq.~\ref{eq:2}) leads to angular-dependent scattering rates $\eta_{\nu}(\theta)$
(Fig.~\ref{fig:fig3}~(c)). At $\theta=0^\circ$ the ratio of scattering rates between the $\gamma$ and
$\beta$ sheets is large, because these bands still have mainly $xy$ and $xz/yz$ character, respectively
(Fig.~\ref{fig:fig2}, inset). As expected from Fig.~\ref{fig:fig3}~(a), this sheet dependence decreases
with increasing temperature. On the other hand, at $\theta=45^\circ$ the ratio is small, due to a very
similar orbital composition of the $\gamma$ and $\beta$ sheets. The $\alpha$ pocket (being almost entirely
$xz/yz$) has the lowest scattering rate and turns $R_H$ positive when $\xi$ becomes large enough at low
temperatures. To shed more light on the interplay of the individual FS sheets we can phenomenologically
assign constant scattering rates to each FS sheet, as shown in Fig.~\ref{fig:fig3}~(d). We see that
for $R_H$ to be positive a necessary condition is $\eta_{\beta} > \eta_{\alpha}$. This again highlights
the importance of SOC, because without SOC the $\alpha$ and $\beta$ sheets have entirely $xz/yz$ orbital
character, and thus $\eta_\alpha = \eta_\beta$. Should one make this assumption also in the presence of
SOC, it would not result in $R_H > 0$ for any ratio $\eta_{\gamma}/\eta_{\alpha}$ (Fig.~\ref{fig:fig3}~(d),
dashed line).

\emph{Impurity-dominated scattering}

\begin{figure}[h]
	\includegraphics[width=1.0\linewidth]{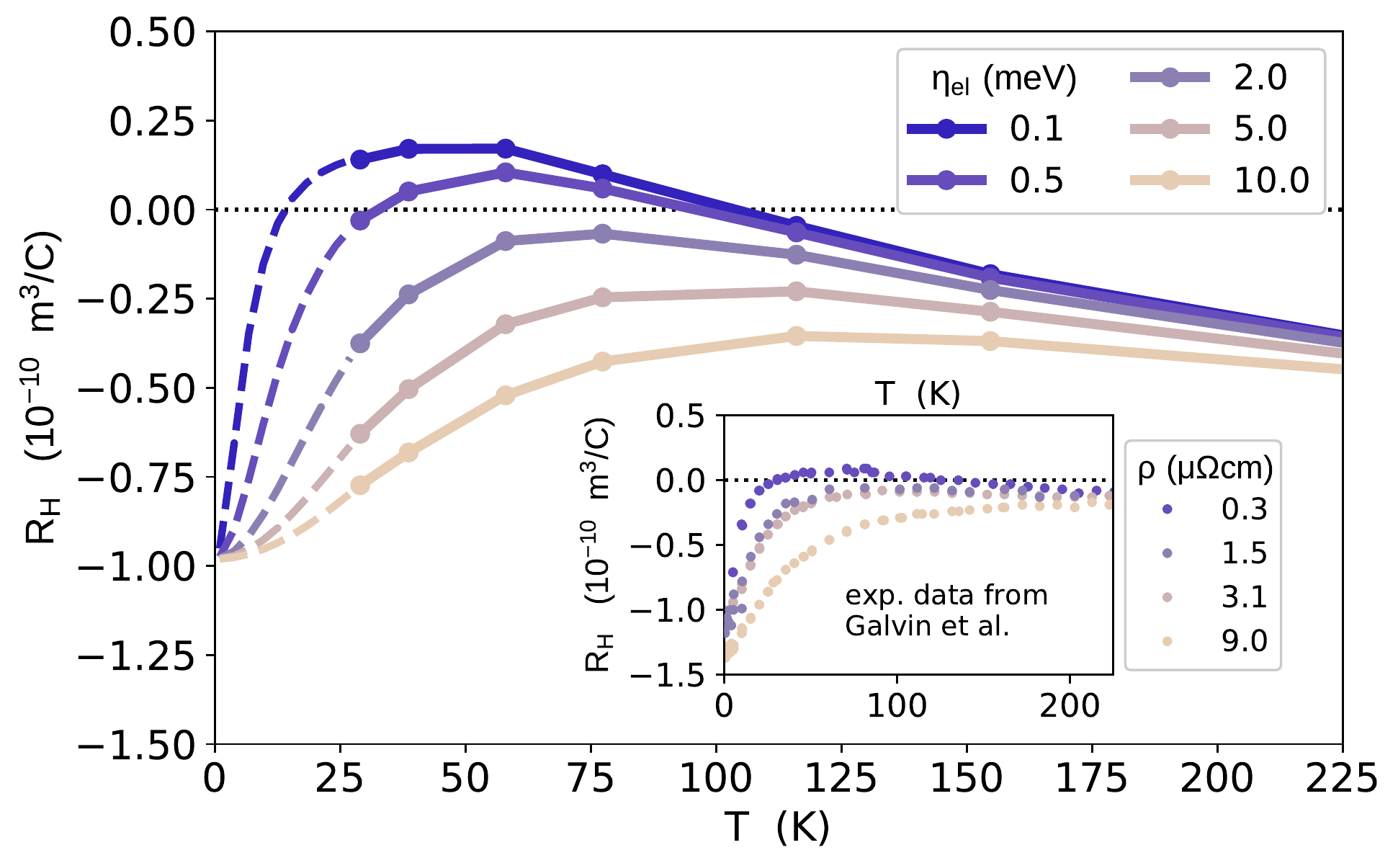}
	\caption{Full temperature dependence of the Hall coefficient $R_H$ with added elastic scattering
		$\eta^{\text{el.}}$ ranging from \num{0.1} to \SI{10}{meV}. The dashed lines indicate the Fermi
		liquid regime with parameters determined from the DMFT result at \SI{29}{K}. The inset shows the
		experimentally measured $R_H$ from Ref.~\cite{Galvin2001} for samples with different residual
		resistivities $\rho$ obtained by introducing small amounts of Al impurities.}
	\label{fig:fig4}
\end{figure}

Considering inelastic scattering only would yield a positive $R_H$ at even lower temperatures deep
in the Fermi liquid regime. However, at such low temperatures elastic scattering is expected to dominate
over inelastic scattering. The extracted DMFT scattering rates at \SI{29}{K} with \SI{5.5}{meV} for the
$xy$ and \SI{1.9}{meV} for the $xz/yz$ orbitals are of the order of the impurity scattering for `clean'
samples with residual resistivities of $\sim\SI{0.5}{\micro\Omega cm}$. Therefore, we add a constant
elastic scattering $\eta^{\text{el.}}$ to the orbital-dependent inelastic scattering $\eta^{\text{inel.}}_{m}$.
This elastic term is assumed to be isotropic: $\eta^{\text{el.}}_{xy}=\eta^{\text{el.}}_{xz/yz}$. The
resulting temperature dependence of $R_H$ for values of $\eta^{\text{el.}}$ ranging from $0.1$ to
\SI{10}{meV} is shown in Fig.~\ref{fig:fig4}. The dashed lines are calculated with the Fermi liquid
form $\eta^{\text{inel.}}_{m} = A_m T^2$ and parameters $A_m$ determined from the calculated inelastic
scattering rates at $\SI{29}{K}$.

For small enough $\eta^{\text{el.}}$ we observe a second zero crossing of $R_H(T)$ and a regime with
$R_H < 0$ at low temperatures, which is consistent with $R_H(T)$ depicted in Fig.~\ref{fig:fig1}.
For $T\rightarrow 0$ the fully elastic scattering regime is reached, and thus $R_H$ is not influenced
by the magnitude of the (isotropic) scattering rate, but rather by the shape of the FS only. This
regime corresponds to $\xi=1$ in Fig.~\ref{fig:fig2}, for which we obtain
$R_H = \SI{-0.94 e-10}{\cubic\meter\per\coulomb}$, in good quantitative agreement with
experiments~\cite{Shirakawa1995,Mackenzie1996,Galvin2001,Kikugawa2004}. With increasing temperature
the influence of elastic scattering fades
away and the precise interplay with inelastic scattering shapes the overall temperature dependence
of $R_H$. Hence, we see that also the low-temperature zero crossing has a simple physical interpretation:
it signals the crossover between the regime dominated by elastic scattering at low temperatures and
the regime dominated by inelastic scattering at higher temperatures. Matching the two terms in the
scattering rate, a simple estimate of the corresponding crossover scale is
$T_1 \sim \sqrt{\eta^{\text{el.}}/A_{xy}} \sim \sqrt{\eta^{\text{el.}}T_{\mathrm{FL}} }$. This scale
obviously depends on the elastic scattering rate, and coincides approximately with the Fermi liquid
coherence scale $T_{\mathrm{FL}}$ only for the cleanest samples reported in which
$\eta^{\text{el.}} \sim T_{\mathrm{FL}}$. For even cleaner samples we predict $T_1 < T_{\mathrm{FL}}$.

On the contrary, for larger $\eta^{\text{el.}}$ we find that $R_H(T)$ ceases to exhibit any zero
crossing and is negative in the whole temperature range. 
Only in very clean samples can the inelastic scattering rate sufficiently exceed the elastic
one for the sign changes of $R_H$ to occur.
This is further substantiated by experimental Hall
measurements for samples where the residual resistivity was altered by introducing different amounts
of Al impurities, cf. the dependence of $R_H$ on $\eta^{\text{el.}}$ in Fig.~\ref{fig:fig4} and the
inset with experimental data from Ref.~\cite{Galvin2001}. 

In the high-T limit we obtain a value of $R_H$ which is more negative than the one reported in
experiments~\cite{Shirakawa1995,Mackenzie1996,Galvin2001}. Within Boltzmann transport theory this
would imply that a larger ratio $\eta_{xy} / \eta_{xz/yz}$ is needed. Likewise, resistivities are
significantly underestimated in DMFT transport calculations for $T>\SI{300}{K}$ in this
material~\cite{Deng2016}. A possible explanation is that other sources of inelastic scattering,
e.g.\ electron-phonon scattering, could play an important role in the high-T regime. We emphasize,
however that all experimental evidence points towards negligible magnetic contribution (due to
processes like skew scattering) and a standard orbital-dominated Hall effect in
\SRO{}~\cite{Mackenzie1996, Noce2000, Kikugawa2004}.

\section*{Discussion}

In summary, our quantitative calculations and qualitative interpretations explain the highly unusual temperature dependence
of the Hall coefficient of \SRO{}. The high-T sign change of $R_H(T)$ in clean samples is the direct consequence of the crossover from a high-T incoherent regime to a coherent
regime with orbital differentiation. 
The orbital composition of each quasiparticle state on the Fermi surface, as well as the
distinct scattering rates of the different orbitals, are crucial
to this phenomenon and are properly captured by DMFT. 
This is in line with recent insights from angle-resolved
photoemission spectroscopy~\cite{Tamai2018}.
In turn, the low-T sign change is due to the crossover from inelastic to impurity-dominated scattering,
which is further substantiated by comparing our results to
experimental data on samples with a higher impurity concentration.
Because it directly affects the shape of the Fermi surface sheets and strongly mixes their orbital character, spin-orbit coupling is found to be essential in explaining $R_H(T)$.

Orbital differentiation is actually a general feature common to Hund's metals~\cite{Aichhorn2010, Mravlje2011, Lanata2013, Miao2016, Medici2017, Kostin2018}, a broad class of materials in which
the electronic correlations are governed by the Hund's coupling, comprising for example
transition metal oxides of the 4d series as well as iron based
superconductors~\cite{Werner2008, Haule2009,  Yin2011, Georges2013, Medici2017}.
We note that a non-monotonic temperature dependence of the Hall coefficient has
also been reported for \ce{Sr3Ru2O7}~\cite{Perry2000}.
Beyond ruthenates, \ce{LiFeAs} and \ce{FeSe} are two compounds without Fermi surface reconstruction
due to long-range magnetic order, which display striking similarities to \SRO{} in many regards. 
The Fermi surface of these superconductors is also
composed of multiple electron- and hole-like sheets with distinct
orbital composition and strong orbital differentiation~\cite{Miao2016, Kostin2018}.
And indeed, the Hall coefficient of \ce{LiFeAs} has a strong temperature dependence~\cite{Heyer2011} and that of \ce{FeSe} displays two sign changes in the tetragonal phase~\cite{Watson2015, Sun2017}.
These examples show that
strongly-correlated materials with multiple Fermi surface sheets of different or mixed orbital character and a orbital-differentiated coherence-to-incoherence crossover are expected to show a pronounced temperature dependence of the Hall coefficient. Sign-changes then emerge in materials with balanced
electron and hole-like contributions. These observations point to a wide relevance of our findings beyond the specific case of \SRO{}.

\clearpage

\clearpage

\clearpage

\clearpage

\clearpage

\section*{Method}

\emph{Hamiltonian and SOC}

We use a maximally-localized Wannier function construction~\cite{MLWF1,MLWF2} to obtain an effective low-energy Hamiltonian for the three $t_{2g}$-like orbitals centered on the \ce{Ru} atoms. This construction
is based on a non-SOC density functional theory calculation (DFT), using the software packages WIEN2K~\cite{Wien2k} with GGA-PBE~\cite{PBE}, wien2wannier~\cite{wien2wannier} and Wannier90~\cite{wannier90}.
We incorporate the SOC as an additional local term, where we neglect the coupling to $e_g$ orbitals, as these are well separated in energy. It has been shown that electronic correlations lead to an effective enhancement of the
SOC in \SRO{} by nearly a factor of 2~\cite{Liu2008,Zhang2016,Kim2018}. As the corresponding off-diagonal elements of the self-energy (in the cubic basis) are approximately frequency-independent~\cite{Kim2018}, we model the effect of correlations by using a static effective SOC strength of $\lambda = \SI{200}{meV}$, instead of the DFT value of about $\lambda^{DFT} = \SI{100}{meV}$. This is crucial to obtain precise agreement with the Fermi surface recently measured with photoemission experiments~\cite{Tamai2018}. We refer to Ref.~\cite{Tamai2018} for further details on the Hamiltonian.

\emph{DMFT}

Due to the fermionic sign-problem, low temperatures are only accessible without SOC.
However, in \SRO{} the diagonal parts of the self-energy are, to a good approximation,
unchanged upon the inclusion of SOC~\cite{Kim2018, Linden2018}.
Therefore, we perform DMFT calculation using the Hamiltonian without SOC and
Hubbard-Kanamori interactions of $U=\SI{2.3}{eV}$ and $J=\SI{0.4}{eV}$
obtained with cRPA~\cite{Mravlje2011}. 
We use the TRIQS library~\cite{TRIQS} in combination
with DFTTools~\cite{TRIQS/DFTTOOLS} and the CT-HYB~\cite{TRIQS,TRIQS/CTHYB}
impurity solver ($\num{3.85e9}$ measurements). 
Inelastic scattering rates are extracted from non-SOC self-energies $\Sigma(i\omega_n)$ by fitting a polynomial of $4^{\text{th}}$ order to the lowest 6 Matsubara points and extrapolating Im$\left[\Sigma\left(i\omega_n \rightarrow 0\right)\right]$, a procedure used in Ref.~\cite{Mravlje2011}. We calculate the standard deviation with 9 consecutive DMFT iterations to obtain errorbars for the inelastic scattering rate ratios.

\emph{Transport calculations}

We calculate $R_H$ using Boltzmann transport theory as implemented in the BoltzTraP package and described in the corresponding Refs.~\cite{Madsen2006, Madsen2018}. We use a $46 \times 46\times 46$ input $\bk$-grid, which is interpolated on a 5 times denser grid with BoltzTraP~\cite{Madsen2006, Madsen2018}.
From Eq.~\ref{eq:2} we obtain a scattering rate for each band and $\bk$-point.
Off-diagonal elements of $\eta_{\nu\nu'}\left(\bk\right)$ and possible
inter-band transitions are not considered in BoltzTraP, but we verified with Kubo transport calculations (TRIQS/DFTTools~\cite{TRIQS/DFTTOOLS}) that these are negligible for the ordinary conductivity $\sigma_{xx}$.
For all transport calculations labeled with `SOC' we use the effective one-particle SOC term with a coupling
strength of $\lambda = \SI{200}{meV}$, however even with $\lambda^{DFT}$ we find the same qualitative
conclusions. We note that recently interest has been devoted to the theoretical descriptions of the Hall effect in strongly-correlated systems beyond Boltzmann transport theory~\cite{Auerbach2018, Nourafkan2018, Miterscherling2018}.

\section*{Data Availability}
All data generated and analyzed during this study are available from
the corresponding author upon reasonable request.

\section*{Acknowledgment}
We gratefully acknowledge useful discussions with Gabriel Kotliar, Andrew Mackenzie, Hugo Strand, \mbox{Andrea}
Damascelli, Reza Nourafkan, Andr\'e-Marie \mbox{Tremblay}. JM is supported by Slovenian Research
Agency (ARRS) under Program P1-0044. MA acknowledges support from the Austrian Science Fund (FWF),
project Y746, and NAWI Graz. This work was supported in part by the European Research Council
grant ERC-319286-QMAC. The Flatiron Institute is a division of the Simons Foundation.

\section*{Author Contribution}
M.Z. performed all calculations and the results were analyzed by M.Z. and A.G. All authors discussed and
interpreted the results at different stages. The whole project was initiated by A.G.
The manuscript was written by M.Z with the help of all authors.

\section*{Additional Information}

\textbf{Competing interests:}
The Authors declare no Competing Financial or Non-Financial Interests

\textbf{Corresponding Authors:}
Manuel Zingl (mzingl@flatironinstitute.org) or Antoine Georges (ageorges@flatironinstitute.org)

\section*{References}
\vspace*{0.5cm}
\bibliography{literatur.bib}
\hspace{0.5in}

\bibliographystyle{naturemag}

\end{document}